\pdfoutput=1

\documentclass[11pt]{article}

\usepackage{ACL2023}

\usepackage{times}
\usepackage{latexsym}
\usepackage{todonotes}
\usepackage{url}
\usepackage{graphicx}
\usepackage{booktabs}
\usepackage{tabularx}
\usepackage{multirow}
\usepackage{colortbl}
\usepackage{xcolor}
\usepackage{amssymb}
\usepackage{amsmath}
\usepackage{makecell}
\usepackage{pgfplots}
\pgfplotsset{width=10cm,compat=1.9}

\usepackage[T1]{fontenc}

\usepackage[utf8]{inputenc}

\usepackage{microtype}

\usepackage{inconsolata}

\usepackage{xcolor}
\definecolor{DarkGreen}{HTML}{108034}
\definecolor{LightGreen}{HTML}{90ee90}
\definecolor{LightYellow}{HTML}{ffee4c}

\newcommand*\rot{\rotatebox{90}}

\newcommand{\openllm}{\textsc{Flexi}}

\newcommand{\DEVELOPMENT}{0} 
\usepackage{ifthen}
\ifthenelse{\DEVELOPMENT = 1}{
	\newcommand{\tz}[1]{\textcolor{red}{\textbf{TZ:} #1}}
    \newcommand{\mh}[1]{\textcolor{blue}{\textbf{MH:} #1}}
	\newcommand{\ns}[1]{\textcolor{blue}{\textbf{NS:} #1}}
 	\newcommand{\dv}[1]{\textcolor{magenta}{\textbf{DV:} #1}}
  \newcommand{\pa}[1]{\textcolor{cyan}{\textbf{PA:} #1}}
}{
\newcommand{\tz}[1]{}	
\newcommand{\mh}[1]{}
\newcommand{\ns}[1]{}
\newcommand{\dv}[1]{}
\newcommand{\pa}[1]{}
}

\newcommand*{\MinNumber}{0.3}%
\newcommand*{\MidNumber}{0.6}%
\newcommand*{\MaxNumber}{0.9}%
\newcommand{\ApplyGradient}[1]{%
        \ifdim #1 pt > \MidNumber pt
            \pgfmathsetmacro{\PercentColor}{max(min(100.0*(#1 - \MidNumber)/(\MaxNumber-\MidNumber),100.0),0.00)} %
            \hspace{-0.33em}\colorbox{LightGreen!\PercentColor!white}{#1}
        \else
            \pgfmathsetmacro{\PercentColor}{max(min(100.0*(\MidNumber - #1)/(\MidNumber-\MinNumber),100.0),0.00)} %
            \hspace{-0.33em}\colorbox{LightYellow!\PercentColor!white}{#1}
        \fi
}

\title{FernUni LLM Experimental Infrastructure (\openllm{}) -- \\Enabling Experimentation and Innovation in Higher Education\\ Through Access to Open Large Language Models}

\author{Torsten Zesch\textsuperscript{1} \and \bf Michael Hanses \textsuperscript{1, 2} \and Niels Seidel\textsuperscript{1} \\ \bf Piush Aggarwal\textsuperscript{1} \and \bf Dirk Veiel\textsuperscript{1} \and Claudia de Witt\textsuperscript{1,2}\\
        \textsuperscript{1}CATALPA, FernUniversität in Hagen, Germany \\ \textsuperscript{2} Institut für Bildungswissenschaft und Medienforschung\\}

\begin{document}
\maketitle
\begin{abstract}
Using the full potential of LLMs in higher education is hindered by challenges with access to LLMs.
The two main access modes currently discussed are
paying for a cloud-based LLM or providing a locally maintained open LLM.
In this paper, we describe the current state of establishing an open LLM infrastructure at FernUniversität in Hagen under the project name \openllm{} (FernUni LLM Experimental Infrastructure).
\openllm{} enables experimentation within teaching and research with the goal of generating strongly needed evidence in favor (or against) the use of locally maintained open LLMs in higher education.
The paper will provide some practical guidance for everyone trying to decide whether to run their own LLM server.
\end{abstract}

\section{Motivation}

While the potential of Large Language Models (LLMs) for higher education has been identified \cite{KASNECI2023102274}, as long as access is not provided by the university in some way, everybody is using the commercial service of their choice, leading to issues including potential data security problems, decreased educational equity, potentially high costs, etc.

\begin{table*}[t]
    \centering
    \small
    \renewcommand{\arraystretch}{1.5}
    \begin{tabularx}{\linewidth}{lcXcX}
    \toprule
              & \multicolumn{2}{c}{\textbf{Closed LLM}} & \multicolumn{2}{c}{\textbf{\textsc{Flexi} Open LLM}} \\
         \midrule
         \bf{Setup costs} 
         & ++ & none 
         & - & server storage; dedicated server\\
         
         \bf{Operating costs} & - - & \makecell[l]{pay per token} & + & \makecell[l]{operating costs of server} \\

         \bf{Maintenance costs} & ++ & \makecell[l]{included in the operating costs} & - - & \makecell[l]{continued maintenance} \\
         
         \bf{Model quality} & ++ & \makecell[l]{access to latest models} & o & only open-weight models \\

         \bf{Model stability} & o & \makecell[l]{might change at any time; little control} & ++ & \makecell[l]{under university control} \\

         \bf{Data protection} & - & \makecell[l]{hard to ensure} & ++ & \makecell[l]{everything stays within own infrastructure} \\

         \bottomrule
    \end{tabularx}
    \caption{Pros and Cons of closed and open LLM provisioning}
    \label{tab:pro-and-con}
\end{table*}

\begin{figure}[t]
\centering
\includegraphics[width=\linewidth]{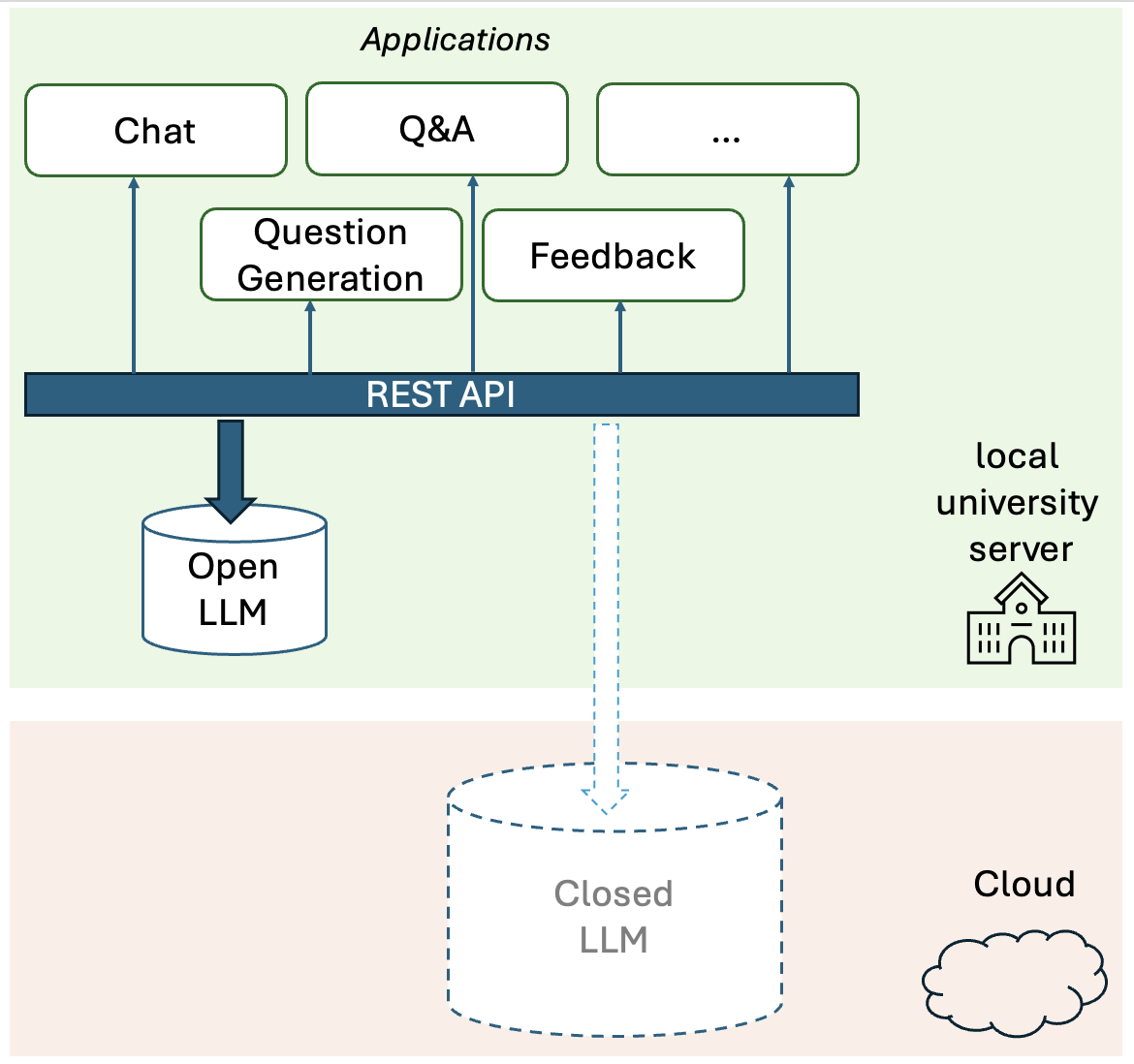}
\caption{The \openllm{} approach replacing a cloud-based LLM with a locally maintained open LLM}
\label{fig:overview}
\end{figure}

Thus, there is an ongoing discussion about whether and how universities should provide LLM access \cite{saldenEtal2024}.
Two main modes are generally being discussed: paying for a cloud-based LLM or providing a locally maintained open LLM.\footnote{Note that there is an ongoing discussion when a model could be called `open' or `open source' \cite{LiesenfeldEtal2024}. We decide to speak of `open models' as soon as the weights are available so we can run them, but we acknowledge (and discuss later in the paper) that different levels of openness come with consequences for academic use.}
Figure~\ref{fig:overview} gives a high-level overview of how a closed, cloud-based LLM could be replaced with an open-source model.
As it shows, replacing a closed LLM with an open LLM can be as easy as pointing the applications to a local REST endpoint once a local LLM is in place.

Both options, commercial and open, have pros and cons, as summarized in Table~\ref{tab:pro-and-con}.
Cloud-based LLMs are easy to set up and run the latest models, while open LLMs are currently more cost-effective overall, and data protection is much easier to achieve.
Based on these considerations, we currently strongly favor open LLMs and decided to implement this setup at FernUniversität in Hagen. 

In this paper, we describe the ongoing efforts at FernUniversität in Hagen to establish such an open LLM infrastructure to enable experimentation in teaching and research.
This research is carried out by researchers from CATALPA (Center of Advanced Technology for Assisted Learning and Predictive Analytics) under the project name \openllm{} (Fernuni LLM Experimental Infrastructure).
The project is supposed to generate strongly needed evidence supporting (or challenging) the use of locally maintained open-source LLMs in higher education settings.
The paper shall give some practical guidance for everyone considering whether to run their own LLM server or not.

Note that the project's scope is currently limited to providing access to text-based, uni-modal LLMs but that other multi-modal services like image generation or speech recognition could be provided using a similar process.

\section{\openllm{} Concept \& Realization}
As we are aiming for an experimental proof-of-concept realization, we aim for a single server that is able to run most open-source models. However, our concept is based on a bare-metal Kubernetes\footnote{\url{https://kubernetes.io}} cluster, which could be extended by additional nodes and thus enable scalable and more robust operation.
We currently ignore guaranteed uptime, redundancy, or other factors that would be central when moving from experimentation to central service delivery.

\subsection{Hardware Setup}
Our concept assumes that the university already operates a data center where the server can be housed. 
Consequently, the university can leverage existing processes for access control, hardware maintenance, network, security, or backup.

The bare-metal Kubernetes setup allows access to all server hardware settings without an additional (extra) virtualization layer. This is very helpful considering the configuration of GPU acceleration.

We first piloted the setup on one server (A) and then replicated it on another server (B). 
Hardware and software specifications of the servers can be found in Table~\ref{tab:specs}.
Both servers were purchased in 2023 for approximately 40,000 €.

\subsection{Software Setup}
For serving the LLMs the open-source project Ollama is used.\footnote{\url{https://ollama.com}}
To ensure optimal performance, GPU acceleration is necessary, but Ollama can also run the models without GPU support. 
NVIDIA and AMD graphic cards are supported. 
Our existing servers have NVIDIA GPU's build in. 
Thus, the combination of operating system, kernel, drivers, and software must match the CUDA version compatibility.\footnote{\url{https://docs.nvidia.com/deploy/cuda-compatibility/}} The operating system is Ubuntu 22.04 LTS on both of our servers with NVIDIA-535-Server and CUDA 12.4 on Server (A) and CUDA 12.2 on Server (B).
On Server (A), both Ollama and Open WebUI\footnote{A webui interface to interact with Ollama models available at \url{https://github.com/open-webui/open-webui}} are deployed as Kubernetes pods. 
On server (B), the docker-compose service is used. For routing and load balancing, the open-source software traefik\footnote{\url{https://traefik.io/traefik/}} is used as ingress-controller \cite{sharma2021traefik}. The usage of containerization allows us to quickly switch between Ollama versions and custom configurations, which is very helpful in this experimental setting.

\begin{table*}[t]
\centering
\small
\renewcommand{\arraystretch}{1.3}
    \begin{tabular}{lll}
    \toprule
    \textbf{Hardware} & \textbf{Server A} & \textbf{Server B} \\
    \midrule
    OS & Ubuntu Server 22.04 LTS & Ubuntu Server 22.04.4 LTS \\
    Kernel & 5.15 & 5.15 \\
    GPU & 8x Nvidia RTX A5000 24GB & 2x Nvidia A40 46GB \\
    CPU & 2x AMD EPYC 7402 &  2x Intel Xeon Gold 6442Y \\
    RAM & 256 GB DDR3 3200 MHZ & 512 GB DDR5 4800 MT/s\\
    Storage & 8 TB SSD &  6 TB SSD \\
    Driver & Nvidia 535 Server &  Nvidia 535 Server \\
    Cuda & 12.4 & 12.2 \\ 
    \bottomrule
    \end{tabular}
\caption{OS and Hardware of \openllm{} Servers}
\label{tab:specs}
\end{table*}

\subsection{Model Selection}
The setup described so far allows us to install and serve any model publicly hosted in the Ollama library.\footnote{\url{https://ollama.com/library}}
At the time of writing, there are over 90 models available, from which we must select a suitable subset. 
Additionally, the web interface enables experimentation with 16,848 models available in GGUF format\footnote{\url{https://github.com/ggerganov/ggml/blob/master/docs/gguf.md}} on Huggingface at the time of writing.




\begin{table}[t]
    \centering
    \small
    \renewcommand{\arraystretch}{1.3}
    \begin{tabular}{lrrr}
    \toprule
     & & \textbf{Context} & \\
    \textbf{Model Name} & \textbf{Size} & \textbf{Length} & \textbf{Licence} \\
    \midrule
    \textsc{Command R+} & 104 B & 128 K & \textit{model specific} \\
    \textsc{DBRX}  & 132 B & 32 K & \textit{model specific} \\
             \textsc{gemma}   &  7 B &  8 K & \textit{model specific} \\
            \textsc{Llama3}  &  8 B &  8 K & \textit{model specific} \\
            \textsc{llava}   & 13 B &  32 K & Apache 2.0 \\
         \textsc{Mistral} &  7 B & 32 K & Apache 2.0 \\
         \textsc{Mixtral} & 22x8 B & 8 K & Apache 2.0 \\

         \textsc{phi3}    &  4 B & 4 K & MIT \\

    \bottomrule
    \end{tabular}
    \caption{Selected LLMs running on the \openllm{} infrastructure (at the time of writing)}
    \label{tab:models}
\end{table}

At the time of writing, we are testing the models listed in Table~\ref{tab:models}.\footnote{While this list is certainly already outdated by the time you are reading this, we still think it serves to give an idea about the range of models being tested.}
We now discuss the dimensions that informed our selection of those models to experiment with.

\subsubsection{Openness}
Open-source large language models (LLMs) come in a variety of `flavors' that significantly differ in how open they actually are \cite{LiesenfeldEtal2024}.
The minimum requirement for our purposes is that the weights are available so we can run, modify, and improve models on our servers. 
\citet{LiesenfeldEtal2024} lists several additional dimensions, including the availability of basic training data that is used for instruction tuning as well as open documentation and a permissive license.

\subsubsection{Language}
Various open models are available that can handle unilingual and multilingual queries. 
For instance, models like \textsc{StableLM2} are trained on multilingual data, including English, Spanish, German, Italian, French, Portuguese, and Dutch. 
The \textsc{Qwen2} model supports 29 languages, including Chinese. 
Specific language models such as \textsc{SauerkrautLM-Qwen-32b} are trained on German language datasets, making them ideal for general-purpose German queries. 
Additionally, the \textsc{Phi3 Medium} model is designed for commercial and research use in English.
However, German language support must be balanced with model quality.

\subsubsection{Quality}

\begin{table*}[t]
    \centering
    \small
    \renewcommand{\arraystretch}{1.3}
    \begin{tabular}{lrccccccc}
    \toprule
    \textbf{Model Name} & [token/s] $\uparrow$ 
        & \rot{\textbf{ARC}} 
        & \rot{\textbf{HellaSwag}} 
        & \rot{\textbf{MMLU}} 
        & \rot{\textbf{TruthfulQA}}
        & \rot{\textbf{WinoGrande}}
        & \rot{\textbf{GSM8K}}  
        & \multicolumn{1}{c}{\textbf{$\varnothing$}}\\
    \midrule


 \textsc{Command R+} & 5 &  
\ApplyGradient{.71} & 
\ApplyGradient{.89} & 
\ApplyGradient{.76} & 
\ApplyGradient{.56} & 
\ApplyGradient{.85} & 
\ApplyGradient{.71} & 
\ApplyGradient{.75} \\

 \textsc{DBRX} & 11 &  
\ApplyGradient{.68} & 
\ApplyGradient{.89} & 
\ApplyGradient{.74} & 
\ApplyGradient{.67} & 
\ApplyGradient{.82} & 
\ApplyGradient{.67} &
\ApplyGradient{.75}  \\
         \textsc{Gemma} & 77  & 	
\ApplyGradient{.65} &
\ApplyGradient{.81} &
\ApplyGradient{.65} &
\ApplyGradient{.55} &
\ApplyGradient{.78} &
\ApplyGradient{.73} &
\ApplyGradient{.69} \\
         \textsc{Llama3} & 82 & 
\ApplyGradient{.73} & 
\ApplyGradient{.86} & 
\ApplyGradient{.80} & 
\ApplyGradient{.64} & 
\ApplyGradient{.83} & 
\ApplyGradient{.88} & 
\ApplyGradient{.79}  \\

         \textsc{Llava} & 60  &  
\ApplyGradient{.53} & 
\ApplyGradient{.76} & 
\ApplyGradient{.52} & 
\ApplyGradient{.46} & 
\ApplyGradient{.72} & 
\ApplyGradient{.15} & 
\ApplyGradient{.52} \\
         \textsc{Mistral} & 94 &  

\ApplyGradient{.73} &
\ApplyGradient{.89} &
\ApplyGradient{.64} &
\ApplyGradient{.78} &
\ApplyGradient{.85} &
\ApplyGradient{.70} &
\ApplyGradient{.77} \\

 \textsc{Mixtral} & 11 &  
\ApplyGradient{.73} & 
\ApplyGradient{.89}  &  
\ApplyGradient{.78}  &  
\ApplyGradient{.68}  &  
\ApplyGradient{.85}  &  
\ApplyGradient{.82}  &  
\ApplyGradient{.79} \\

         \textsc{Phi3} & 127 &  
\ApplyGradient{.67} & 
\ApplyGradient{.86} & 
\ApplyGradient{.78} & 
\ApplyGradient{.58} & 
\ApplyGradient{.73} & 
\ApplyGradient{.80} & 
\ApplyGradient{.74} \\


    \bottomrule
    \end{tabular}
    \caption{LLM model quality based on established benchmarks covering different application areas. Normalized scores range from 0-1. Higher scores correspond to better models. For comparison, we also list the throughput of the models on our Server B in tokens per second.}
    \label{tab:modelperformance}
\end{table*}

A wide range of benchmarks are available on which LLM model quality can be evaluated.
Benchmarks can have different specializations, e.g.\ focusing on language capabilities, world knowledge, common sense reasoning, or coding.
We argue that some capabilities are more important in our educational settings than others.
For example, medical knowledge might not be central at FernUniversität, while knowledge of German \cite{pfister-hotho-2024-supergleber} or factual correctness seems more important.
We discuss here our selection of benchmarks:

\begin{description}
    \item[ARC] \cite{clark2018think} examine LLMs on 7,787 grade-school science questions. The test is challenging and demands extensive general knowledge and strong reasoning skills. It includes two sets: Easy and Challenge (with particularly difficult tasks). 
    \item[GSM8K] \cite{cobbe2021training} is a set of 8,500 grade-school math problems, each requiring two to eight steps to solve using basic math operations. The questions are simple enough for a smart middle schooler to solve and are useful for testing LLMs’ ability to handle multistep math problems.
    \item[HellaSwag] \cite{zellers2019hellaswag} This benchmark evaluates natural language inference (NLI) by prompting LLMs to complete a given passage. What adds to its difficulty is using adversarial filtering to create deceptive yet plausible incorrect answers for the tasks.
    \item[MMLU] stands for Massive Multitask Language Understanding \cite{hendrycks2020measuring} measures general knowledge across 57 different subject areas, spanning from STEM to social sciences. The difficulty levels range from elementary to advanced professional.
    \item[TruthfulQA] \cite{lin2022truthfulqa} aims to determine if LLMs produce incorrect answers based on common misconceptions. The questions cover various categories, including health, law, fiction, and politics.
    \item[WinoGrande] \cite{sakaguchi2021winogrande} is a massive set of 44,000 problems derived from the Winograd Schema Challenge \cite{levesque2012winograd}. These problems consist of nearly identical sentence pairs with two possible answers, where the correct answer depends on a trigger word. This tests the ability of LLMs to accurately understand context. 
\end{description}

\noindent
Table~\ref{tab:modelperformance} illustrates the performance based on the Huggingface leaderboard.\footnote{\url{https://huggingface.co/spaces/open-llm-leaderboard/open_llm_leaderboard}}
The evaluation score has been normalized for each benchmark between 0 and 1, with higher scores indicating better performance.
With the exception of \textsc{Llava}\footnote{\textsc{Llava} is focused on visual tasks, which our selection of benchmarks does not reflect.}, most models perform well on average but have their specializations.
For example, \textsc{Mistral} is especially good on TruthflQA but not on Math problems (GSM8K), whereas \textsc{Llama3} is best.

We also show in the table throughput\footnote{measured via
\url{https://github.com/aidatatools/ollama-benchmark}}, as model quality has to be balanced with how fast the requests can be served.
Combining these two metrics enables us to select the model best suited for a specific use case in \openllm{}.

\subsubsection{Safety}
Applications of LLMs within higher education raise concerns about their security and potential vulnerabilities. 
Ensuring LLM security involves preventing misuse by malicious actors or avoiding unintentional errors, such as accidentally revealing email addresses. Unlike traditional cybersecurity, LLM security depends significantly on natural language processing (NLP) techniques because most attack strategies are language-based. Attacks can occur due to conflicts between application builders, end-users, and external tool outputs, especially when there is explicit knowledge about the builder's intentions or policies \cite{wei2024jailbroken}. Therefore, it is crucial for a secure model to undergo various vulnerability assessments before being deployed. LLM security evaluation frameworks, such as the `Generative AI Red-teaming and Assessment Kit' (garak), facilitate this process \cite{garak}. Through systematic probing, it helps users identify vulnerabilities in language models or dialog systems.
Checks such as profanity, toxicity, encoding flaws, and jailbreaks are analyzed to evaluate model safety. 
The leaderboard scores are available\footnote{\url{https://huggingface.co/spaces/AI-Secure/llm-trustworthy-leaderboard}}, that can estimate these checks for models compatible with \openllm{}.

\subsubsection{Size}
Finally, as \openllm{} operates with the limited resources of a public university, model size is an issue, as bigger models might not run at all on our hardware, or throughput might be insufficient.
As our main goal here is experimentation, we include a few models from the fringes of the size distribution but mainly focus on mid-distribution models.
However, most currently available open models would run on our servers, but throughput and maximum concurrent queries might be insufficient (see sections~\ref{sec:experiences} and \ref{sec:scaling}).

\subsection{Maintenance \& Monitoring}

We use Checkmk\footnote{\url{https://checkmk.com}} to monitor the servers (A) and (B) and the resources they contain.
The so-called Checkmk agent runs on our servers, which collects data from the local system via plugins and transmits it to the backend. 
The backend receives and manages the data and makes it available via dashboards.
Our data center (Zentrum für Digitalisierung und IT, ZDI) operates the backend.

\begin{figure*}[t]
\centering
\includegraphics[width=\linewidth]{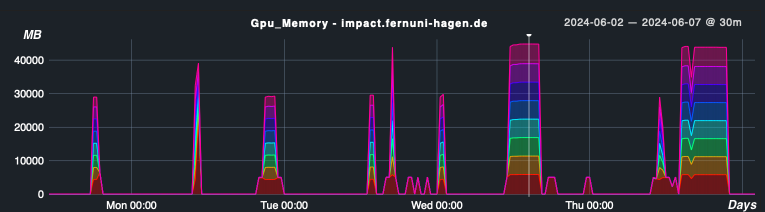}
\caption{Usage spikes during a longer period of use of Server (A). The memory usage of all 8 GPUs is shown in color. You can see that usually, all 8 GPUs are used. However, as soon as it is sufficient to use only one GPU, e.g., for a smaller model (see smaller red spikes), the system implements this accordingly and in a resource-saving manner.}
\label{fig:dashboard}
\end{figure*}

To monitor GPU utilization in particular, we use a special script as a local plugin. The script captures the GPU data (via nvidia\_smi) and passes it to the backend so that the visualizations shown in Figure~\ref{fig:dashboard} can be viewed there. Using the dashboard, we can see the current utilization of the servers, especially the GPUs, and the trend over the past days and weeks. This allows us to identify both peak loads and average server utilization. The knowledge gained in this way is used to better determine the configuration of the servers for regular operation.

\subsection{Data Protection}
When using a cloud-based LLM, all requests are sent to the cloud provider, enabling user tracking and possibly exposing sensitive data.
Projects like HAWKI\footnote{\url{https://github.com/HAWK-Digital-Environments/HAWKI}} solve the tracking issue by bundling all requests from one university so that chats cannot be attributed to a specific person.
Commercial providers like Microsoft offer services like `Azure OpenAI' where the data is not sent to OpenAI (but to Microsoft), and the requests are guaranteed not to be used for model training.
One key advantage of \openllm{} is that requests never leave the premises of the university IT infrastructure (cf.\ Figure~\ref{fig:overview}).

\section{Experiences}
\label{sec:experiences}

In this section, we describe the most important experiences and takeaways from experimenting with \openllm{}.

\paragraph{Load Test}

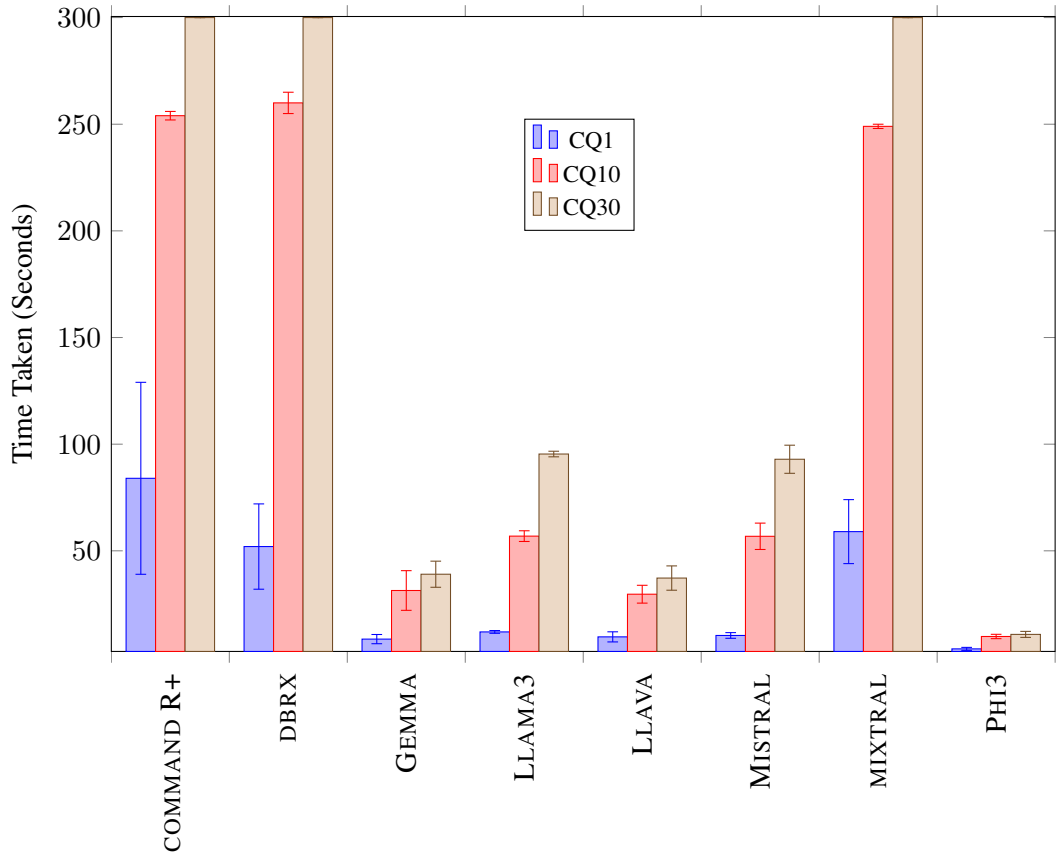
\begin{figure*}[t]
\centering
\begin{tikzpicture}
\begin{axis}[
width=14cm,
height=10cm,
legend style={at={(axis cs:4.0,200)},anchor=south west, font=\small},
enlargelimits={abs=0.5},
ybar=0pt,
bar width=0.25,
xticklabel style = {rotate=90,anchor=east},
xtick={0.5,1.5,...,10.5},
xticklabels={\textsc{command R+}, \textsc{dbrx}, \textsc{Gemma}, \textsc{Llama3}, \textsc{Llava}, \textsc{Mistral}, \textsc{mixtral}, \textsc{Phi3}},
x tick label as interval,
ylabel=Time Taken (Seconds),
]
\addplot+[error bars/.cd,
y dir=both,y explicit]
coordinates {
(1,84) +- (0.0, 45)
(2,52) +- (0.0, 20)
(3,8.61) +- (0.0, 2.16)
(4,11.98) +- (0.0, 0.65)
(5,9.66) +- (0.0, 2.37)
(6,10.32) +- (0.0, 1.33)
(7,59) +- (0.0, 15)
(8,4.03) +- (0.0, 0.7)

};
\addplot+[error bars/.cd,
y dir=both,y explicit]
coordinates {
(1,254) +- (0.0, 2)
(2,260) +- (0.0, 5)
(3,31.4) +- (0.0, 9.3)
(4,56.89) +- (0.0, 2.51)
(5,29.66) +- (0.0, 4.18)
(6,56.81) +- (0.0, 6.18)
(7,249) +- (0.0, 1)
(8,9.84) +- (0.0, 1.02)
};
\addplot+[error bars/.cd,
y dir=both,y explicit]
coordinates {
(1,300) +- (0.0, 0)
(2,300) +- (0.0, 0)
 (3,39.03) +- (0.0, 6.12)
(4,95.38) +- (0.0, 1.31)
(5,37.23) +- (0.0, 5.7)
(6,92.93) +- (0.0, 6.59)
(7,300) +- (0.0, 0)
(8,10.82) +- (0.0, 1.41)
};

\legend{CQ1, CQ10, CQ30}
\end{axis}
\end{tikzpicture}
\caption{Load test results accessing Server B. We show results for a single query as well as 10 and 30 concurrent queries (CQ). There is a response timeout of 300s, so this is the maximum possible average.}
\label{loadtest_landscape}
\end{figure*}

To analyze this, we attempted to test our server's load using a general-purpose laptop. 
We sent multiple REST API POST requests to \openllm{}. 
To fully utilize the available GPU space, multiple models are initiated simultaneously. This approach will make efficient use of the GPUs and significantly reduce the server response time. 
In addition, multiple instances of smaller models can be created, allowing it to handle multiple requests simultaneously.
Figure~\ref{loadtest_landscape} illustrates the load in terms of time taken by different models on \textsc{Server (B)} while handling concurrent requests. 
On the one hand, models such as \textsc{phi3} can handle multiple requests with hardly an increase in latency time. On the other hand, models such as \textsc{mistral} and \textsc{llama3} exhibit higher latency with increased concurrent queries. 
For synchronous tasks like ChatBots, the response times of larger models will probably be too high once this is scaled to many users.
However, not all tasks require the largest models, especially since benchmark quality improvements are often marginal (cf.\ Figure~\ref{tab:modelperformance}).

\paragraph{Operating Costs}
Assuming that the data center itself already has fixed costs for the university, operating costs are dominated by energy demand\footnote{FernUniversität in Hagen is already using photovoltaic systems to meet some of the university's energy needs.}.
In contrast to closed LLM servers, where very little information about energy usage is available, we can directly measure energy usage.
For the 5-day period shown in Figure~\ref{fig:dashboard}, approximately 26.7 kilowatt-hours (kWh) were consumed by 8 GPUs, i.e.\ about 5 kWh per day.
The theoretical maximum, which we have not measured yet, would be around 44.16 kWh a day or 16 MWh a year.
At 0.30 € per kWh, this translates to a maximal annual operating cost for the 8 GPUs of about 5,000~€.

This energy usage translates into 6 tons of emitted CO$_2$.\footnote{Germany 2023: 380 gCO2 per kWh, \url{https://ember-climate.org/data/data-tools/data-explorer/}}
However, this number has to be put into perspective, as recent research \cite{tomlinson_carbon_2024} indicates that LLM energy usage is dwarfed by the energy usage of the computers being used to access the LLMs (e.g.\ by students using Moodle to access their course contents).

\section{Applications \& Use Cases}
\label{sec:use-cases}

While a setup as implemented through \openllm{} can support a wide range of applications \cite{Rashid2024}, we list here some use cases that we believe to be of specific interest in higher education.

\begin{figure*}[t]
\centering
\includegraphics[width=\linewidth]{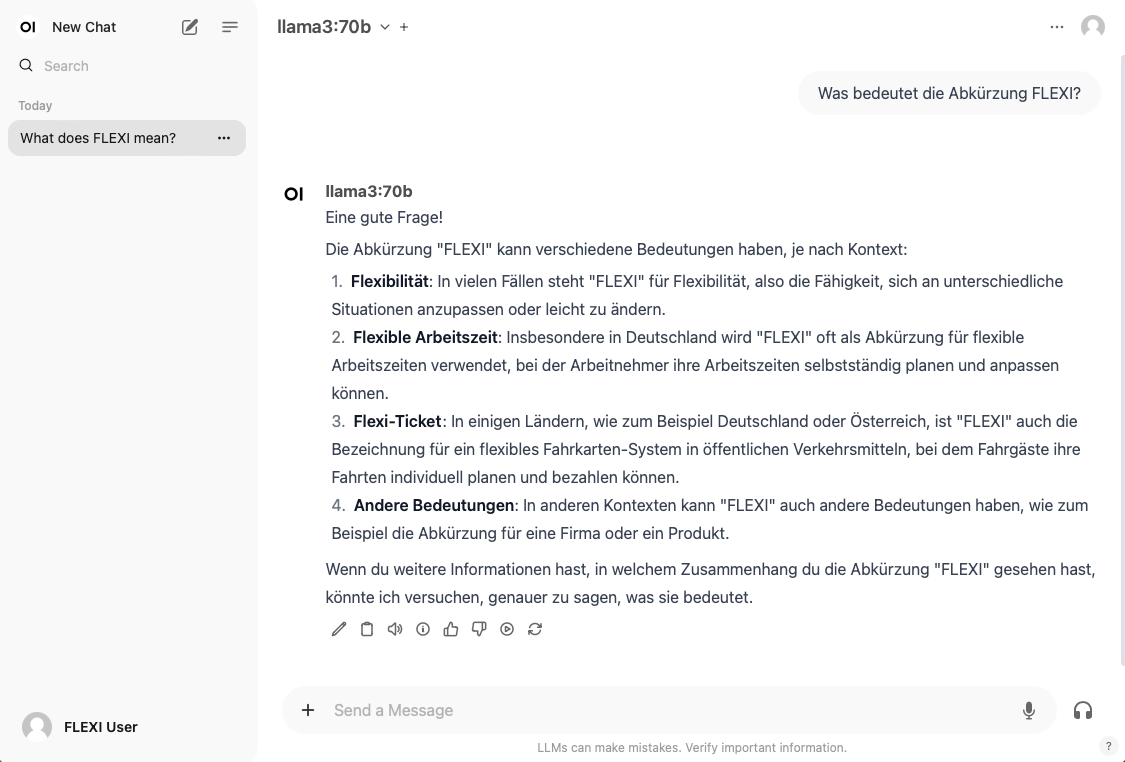}
\caption{Screenshot of an example session with the OpenWebUI chat frontend using the Llamma3-70b model.}
\label{fig:openwebui}
\end{figure*}

\paragraph{Chat Interface}
To assist the students and educators, as an entry point familiar to everyone who has used the web interface of, e.g.,\ ChatGPT, we provide a chat interface based on OpenWebUI.\footnote{\url{https://github.com/open-webui/open-webui}}
Figure~\ref{fig:openwebui} shows an example chat session.

\paragraph{RAG}
We are experimenting with retrieval-augmented generation (RAG) applications, where, e.g., \ lecture notes are indexed, and students are provided with access to a dedicated chatbot that can answer questions regarding the study material. 

\paragraph{API Access} Selected users who want to work with the API are granted direct API access and may implement their own applications.
For example, the \url{https://what2study.de} project uses direct API access to experiment with their system.

\paragraph{LMS Integration}
We are developing middleware for LLM access in Moodle, a widely used open-source learning management system (LMS), under the name \textsc{Caipi} (CATALPA AI Prompting Interface).
At FernUniversität in Hagen, numerous students and lecturers use Moodle to access courses and interact with learning materials and each other.
Integrating LLM access into Moodle is crucial for enhancing these interactions.

There already are Moodle plugins making use of LLMs directly, for use cases like question generation\footnote{e.g.\\ \url{https://moodle.org/plugins/local_aiquestions},\\ \url{https://moodle.org/plugins/block_openai_questions}.} or as chatbots.\footnote{e.g.\\
\url{https://moodle.org/plugins/block_openai_chat},\\
\url{https://moodle.org/plugins/local_ulibot},\\
\url{https://moodle.org/plugins/block_ube_ta},\\
\url{https://moodle.org/plugins/block_openai_chat}}
They all implement LLM access directly.
\textsc{Caipi} creates an abstraction layer, establishing an authorized interface between user requests and the API acces provided by \openllm{}.
\textsc{Caipi} structures requests, checks input parameters, regulates access based on user roles, and enables load balancing.
Prompts, including parameters from the Moodle database, can be stored to be re-used.
\tz{mögliches Datenschutzproblem? lieber weglassen hier?}


\section{Future work: University-wide scaling}
\label{sec:scaling}

\openllm{} is an experiment aimed at learning more about the possible pitfalls of providing open LLM access.
Access is thus currently limited to selected early adopters who know about models' possible shortcomings, who do not expect flawless operation, and who are giving us valuable feedback on how to improve the service.

Should we eventually want to drop the `experimental' status and provide the same service on a university-wide level, we have some more challenges ahead of us.
We would probably need a bigger \textbf{server} and an \textbf{operational concept} providing guaranteed uptime, redundancy, and load balancing.
We would also have to look deeper into \textbf{legal issues} which we discuss here for the specific situation at our university in Germany: 

First, the university's IT administration requires an operational concept for the regular operation of such a service. 
This includes the system architecture, security measures, maintenance routines, and responsibilities.
Second, compliance requirements regarding the General Data Protection Regulation (GDPR) must be met. 
\tz{die Referenz funktioniert so schlecht, weil komischer Autor und das "and" im Namen wird als Autorentrenner interpretiert.}
This includes a document titled ``record of processing activities'' detailing how personal data (e.g.\ server logs, access logs, and possibly user inputs) are processed. 
It specifically describes users' rights regarding information access, data retention periods, and deletion of personal data.
Third, a user agreement is necessary, which users must accept upon their first access to an \openllm{}-based application.
This agreement includes information on the processing of personal data and disclaimers about the reliability of information generated by the LLMs.

Beyond legal issues, we are also facing \textbf{ethical questions}.
Which level of freedom of speech should a university allow?
The answer might differ depending on the use case, where research probably needs less censored models than teaching.
Universities also may need to take a stance on which political orientation an LLM should express, as they have been found to vary quite a lot \cite{feng-etal-2023-pretraining}.
However, recent research \cite{röttger2024political} somewhat challenges this view, finding that ``even small changes in situative context can substantially affect the values and opinions manifested in LLMs''.
Consequently, any decisions on which LLM is suitable in a higher education context should only be made if tested given salient use cases.
It is the role of experiments like \openllm{} to provide an environment where these kinds of tests can be conducted.

\section{Summary}
In this paper, we have argued that in an academic context, locally hosting open-source LLMs is currently the best choice, balancing the pros (data protection and cost) with the cons (maintenance effort).
To that end, we have set up \openllm{}, a concrete implementation example that can serve as a reference point for others setting out on a similar endeavor.
With moderate hardware, we have shown that it is possible to provide access to open LLM and thus support a wide range of educational applications.
\openllm{} provides maximal data protection, as no LLM request ever leaves the premises of our university.
While our approach was designed and implemented for the higher education sector, it may be applied to other sectors or domains as well.
A key open question remains model selection, as a wide variety is on offer and use cases might have different needs.
A solution could be to run multiple models in parallel, as we are successfully doing right now.

\section*{Acknowledgments}
This research was supported by the Center of Advanced Technology for Assisted Learning and Predictive Analytics (CATALPA) of FernUniversit\"at in Hagen, Germany.
Enabling access to LLMs at FernUniversit\"at is a central activity organized by Zentrum für Lernen und Innovation (ZLI).

\bibliography{custom}
\bibliographystyle{acl_natbib}

\appendix
\onecolumn
\clearpage
\section{Server Load Results}
\label{finegrainedload}
\begin{table*}[h!]
    \centering
    \small
    \begin{tabular}{lrrr|rrr|rrrr}
    \toprule
    & \multicolumn{9}{c}{Time [s]} \\
    \textbf{Model Name (Size)} & 
    \multicolumn{3}{c}{\textbf{1 CQ}} & \multicolumn{3}{c}{\textbf{10 CQ}} & \multicolumn{3}{c}{\textbf{30 CQ}}  \\
    \midrule
     \multicolumn{11}{l}{\textbf{Prompt:} Write a step-by-step guide on how to bake a chocolate cake from scratch.} \\
     \addlinespace[1mm]
     \textsc{command R+} & 106 &± & 20  &  300&± &0    & 300&± &0 \\
     \textsc{dbrx} &  61 & ± & 7  &  300&± &0    & 300&± &0  \\
              \textsc{gemma} & 10 &± & 3  &  35 &±& 4    &  53&± &42 \\
             \textsc{Llama3}  &  14  &±  & 1  &  77  &±  & 2  &  127   &±   &5 \\
             \textsc{llava}  &  8 &± &3   &  41 &±& 18 &  47 &±& 13\\
          \textsc{Mistral}  &  11& ± &1  &  66& ± &11    &  104& ± &7\\
\textsc{mixtral}& 76 & ± & 12 &  300&± &0    & 300&± &0 \\
          \textsc{phi3}    & 4& ± &3  &  12& ± &5  &  11& ± &4 \\

         \addlinespace[3mm]
     \multicolumn{11}{l}{\textbf{Prompt:} Develop a python function that solves the following problem,  sudoku game} \\
          \addlinespace[1mm]

          \textsc{command R+} & 110 &± & 69  &  300&± &0    & 300&± &0 \\
     \textsc{dbrx}& 66 & ± & 12  &  300&± &0    & 300&± &0 \\
              \textsc{gemma}&  9 & ± &8  &  23& ± &6  &  28& ± &12 \\
          \textsc{Llama3}  &  14& ± &2 &  68& ± &4 &  120& ± &3\\
         \textsc{llava} &  14& ± &15 &  26& ± &9    &  39& ± &10 \\

        \textsc{Mistral} &  12& ± &1 & 86& ± &15   &  144& ± &12 \\
        \textsc{mixtral} & 83 & ± & 48  &  300&± &0    & 300&± &0 \\
                \textsc{phi3}   &  5& ± &3   &  13& ± &4  &  14& ± &3\\

         \addlinespace[3mm]
     \multicolumn{11}{l}{\textbf{Prompt:} Create a dialogue between two characters that discusses economic crisis} \\
          \addlinespace[1mm]

          \textsc{command R+} & 101 &± & 58  &  300&± &0    & 300&± &0  \\
     \textsc{dbrx} & 46 & ± & 7  &  300&± &0    & 300&± &0  \\
                   \textsc{gemma} &  10&±&8   & 42&±&13  &  63&±&9\\
              \textsc{Llama3} &  13&±&1 &  64&±&4   &  103&±&6 \\
             \textsc{llava} & 6&±&1  &  28&±&11   &  30&±&10\\
          \textsc{Mistral} & 10&±&1 &  45&±&8   &  72&±&9
          \\

        \textsc{mixtral}& 59 & ± & 4  &  300&± &0    & 300&± &0 \\
        \textsc{phi3}   & 3&±&1   &  7&±&1  &  8&±&1\\

         \addlinespace[3mm]
     \multicolumn{11}{l}{\textbf{Prompt:} In a forest, there are brave lions living there.  Please continue the story.} \\
          \addlinespace[1mm]

          \textsc{command R+}& 80 &± & 71  &  300&± &0    & 300&± &0  \\
     \textsc{dbrx} & 41 & ± & 10  &  300&± &0    & 300&± &0 \\

              \textsc{gemma} &  10&± &3  &  33&± &7   &  44&± &7\\
         \textsc{Llama3} &  13&± &9   &  52&± &7  &  92&± &6\\

          \textsc{llava} &  13&± &9   &  41&± &15    &  56&± &17\\

                   \textsc{Mistral} & 11&± &5  &  58&± &19  &  97&± &3\\

        \textsc{mixtral} & 56 & ± & 12 &  300&± &0    & 300&± &0 \\
        \textsc{phi3}   &  4&± &1   &  10&± &3   &  10&± &3\\

         \addlinespace[3mm]
     \multicolumn{11}{l}{\textbf{Prompt:} I'd like to book a flight for 4 to Seattle in U.S.}  \\     
     \addlinespace[1mm]

          \textsc{command R+} & 21 &± & 8  &  69 &±& 8    &  300&± &0 \\
     \textsc{dbrx} & 26 & ± & 4  &  100 &±& 28    &  300&± &0 \\

                  \textsc{gemma} &  5& ± &1  &  10& ± &1    &  11& ± &4\\
                  \textsc{Llama3} &  8& ± &1  &  23& ± &	3  &  35& ± &3 \\

         \textsc{llava} &  7& ± &5  &  8& ± &0   &  12& ± &1\\
             \textsc{Mistral} &  7& ± &1  &  29& ± &7  &  48& ± &6\\

        \textsc{mixtral} & 21 & ± & 0  &  43 &±& 6    &  300&± &0 \\
            \textsc{phi3}   &  4& ± &2  &  9& ± &1   &  11& ± &1\\







    \bottomrule
    \end{tabular}
    \caption{Concurrency test on Server B using multiple concurrent queries (CQ). There is a response timeout of 300s, so this is the maximum possible average.}
    \label{tab:concurrencytest}
\end{table*}



\end{document}